\documentclass[12pt,preprint]{aastex}

\newcommand{\spc}{.6cm}

\shorttitle{Magneto-Turbulent Entrainment}
\shortauthors{Watson et al.}
\received{Yesterday}
\begin{document}
\title{Kelvin-Helmholtz Instability in a Weakly Ionized Medium}
\author{C. Watson\altaffilmark{1}, E.G. Zweibel\altaffilmark{1,2}, 
  F. Heitsch\altaffilmark{1,3}, E. Churchwell \altaffilmark{1}}
\altaffiltext{1}{Univ. of Wisconsin - Madison, Dept. of Astronomy, 475
N. Charter St., Madison, WI 53716}
\altaffiltext{2}{Center for Magnetic Self Organization in Laboratory
  \& Astrophysical Plasmas, Univ. of Wisconsin, Madison}
\altaffiltext{3}{Institute for Astronomy \& Astrophysics, Scheinerstr. 1, 81679 Munich,Germany}
\begin{abstract}
Ambient interstellar material may become entrained in outflows from massive
stars as a result of shear flow instabilities. We study the linear theory
of the Kelvin - Helmholtz instability, the simplest example of shear flow
instability,
in a partially ionized medium. We model the interaction
as a two fluid system (charged and neutral) in a planar geometry. Our
principal result is that for much of the relevant parameter space,
neutrals and ions are sufficiently decoupled that the neutrals are
unstable while the ions are held in place by the magnetic field. Thus,
we predict that there should be a detectably narrower line profile in
ionized species tracing the outflow compared with neutral species
since ionized species are not participating in the turbulent interface
with the ambient ISM. Since the magnetic field is frozen to the plasma, it
is not tangled by the turbulence in the boundary layer.
\end{abstract}
\section{Introduction}
In the process of forming, massive ($>$ 10 M$_\odot$) stars produce
massive bipolar outflows. In contrast to outflows from low-mass
proto-stars, outflows from massive stars frequently have more mass
than their presumed driving proto-star (Churchwell 1997). The
canonical explanation for the origin of the mass is that a less
massive jet originating from either the accretion disk or protostar
sweeps up and entrains ambient interstellar material. This model can
potentially explain both the large masses and poor collimation of
massive outflows. It may also explain how low-mass outflows become
less collimated as they age (Bachiller and Tafalla,
1998). Churchwell~(1997) argued that mechanical entrainment, through
e.g. bow-shocks, could not simultaneously explain both the large mass
and young age of massive outflows.

Shear flow, or Kelvin-Helmholtz, instabilities at the interface
between the jet and the ambient material are an alternative possible
mechanism for entrainment. The instability amplifies ripples at the
interface, culminating in the development of a turbulent boundary
layer which can transfer momentum from the jet to the ambient
medium. A magnetic field parallel to the flow tends to suppress the
instability because magnetic tension opposes rippling. The result is
that instability is only present if the velocity shear is greater than
the Alfv\'en velocity (Chandrasekhar, 1961).

The nonlinear development of the Kelvin-Helmholtz instability is
strongly influenced by the magnetic field (Malagoli et al., 1996 \&
Jones et al., 1997). The numerical simulations performed by these
authors show that the vortices created by the instability wind up the
field until it reaches the Ohmic scale. Windup transfers energy to
small scales, while Ohmic dissipation heats the layer. These effects
are absent in the field-free case.

In the following study, we investigate the Kelvin-Helmholtz
instability in the linear, partially ionized regime to determine its
possible contribution to entrainment in massive bipolar outflows. This
problem cannot be analyzed in the hydrodynamic regime because the
neutral Alfv\'en velocity (v$_A = \frac{B}{\sqrt{4\pi\rho_n}}$) is
sometimes larger than the flow speed, indicating that the magnetic
field can be strong enough to affect the dynamics. On the other hand,
the pure magnetohydrodynamic (MHD) regime is inappropriate because at
sufficiently short wavelengths, Alfv\'en waves propagate nearly
independently of the neutrals.

In \S 2 we describe a model that we believe is appropriate for massive
outflows driven by massive protostars. In \S 3 we discuss the
numerically-derived solutions to the dimensionless dispersion equation
and equivalent analytical solutions to simplified versions of the
dispersion equation. We demonstrate that there is a new instability,
unique to the partially ionized regime, which results from ion-neutral
slip.  In \S 4, we discuss the observational implications of our
results, concentrating on the physical environment of the interaction
zone between a bipolar outflow and the ambient interstellar material
(ISM). \S 5 is a critical summary of the paper.

\section{Model Description}

 We model the interaction zone between a high
speed jet and the ambient ISM by a slab geometry with a flow along the $
x$-axis. The interface between the two fluids is the $x-y$ plane. The
density is everywhere uniform, and the velocity is uniform except at
$z=0$, where it has a step. We assume the flow velocity $U$ decreases
with increasing $z$. The magnetic field is uniform and aligned with the
flow, which maximizes its stabilizing effect in the $x$-direction.

In this model, the equilibrium flow has zero vorticity except on the
plane $z=0$, where the vorticity is infinite. For our choice of
coordinates and velocity profile, 
the total vorticity integrated across the sheet is 
$-2U\hat y$.

\begin{figure}
\plotone{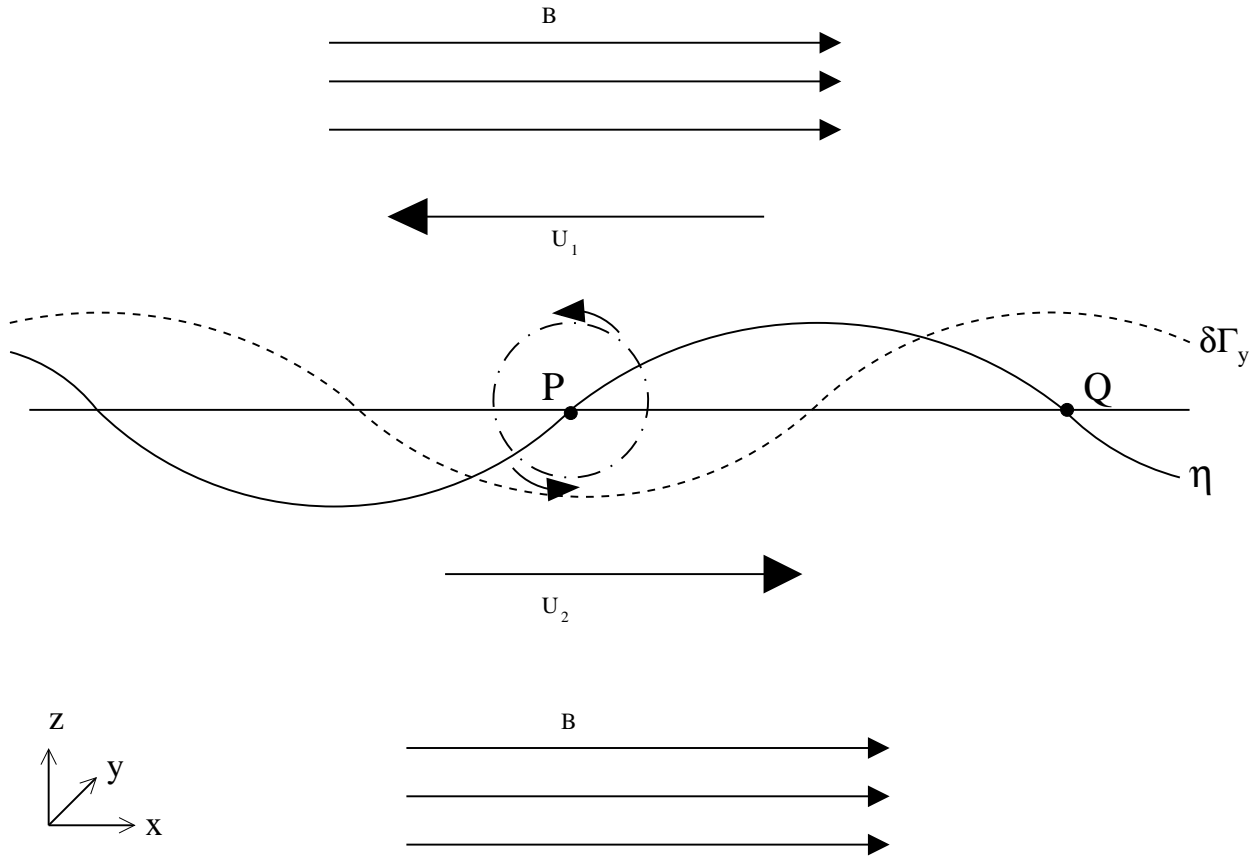}
\caption{Diagram of the standard Kelvin-Helmholtz instability. Since
  the fluid is assumed to be incompressible, a small periodic
  perturbation at a shear layer will tend to grow. In general, the
  vorticity perturbation ($\delta\Gamma_y$) is out of phase with the
  vertical displacement of the vortex sheet ($\eta$). If it can be
  arranged that the vorticity has the distribution shown by the dashed
  line, then the equilibrium flow sweeps negative vorticity toward
  point $P$. The effect is as if an ``extra eddy", rotating
  counterclockwise were added to the flow. The centrifugal force
  associated with the eddy pushes the interface between $P$ and $Q$
  outward, resulting in the growth of $\eta$ with
  time. \label{khdiagram}}
\end{figure}

The mechanism of the Kelvin-Helmholtz instability is explained in
Batchelor (1967). Suppose the interface is perturbed by a small
amplitude ripple, as shown, in the center of mass frame, in
Figure~\ref{khdiagram}. Denote the vertical displacement of the vortex
sheet by $\eta$. In general, the vorticity perturbation is out of
phase with $\eta$. If it can be arranged that the vorticity has the
distribution shown by the dashed line, then the equilibrium flow
sweeps negative vorticity toward point $P$ on both
sides of the interface. The effect is as if an
extra eddy, rotating counterclockwise as sketched in the Figure,
were added to the flow. The centrifugal force associated with the eddy
pushes the interface between $P$ and $Q$ outward, resulting in the
growth of $\eta$ with time. If the vorticity perturbation were shifted
in phase by $\pi$ with respect to the dashed line, $\eta$ would
shrink. As explained here, the instability mechanism requires
advection of vorticity by the bulk
flow in opposite directions above and below the
interface. It is always possible to choose a reference frame in which
this is the case, provided that the velocity profile has a jump.

A magnetic field parallel to the flow exerts a tension force when
bent. The tension force is stabilizing, and the criterion that it
dominates hydrodynamic forces is roughly that the magnetic energy
exceeds the kinetic energy of relative motion, or, equivalently, that the
Alfv\'en speed for the bulk medium exceed the flow speed.

The basic mechanism of instability is unaffected by partial
ionization. In the weakly ionized medium of interest here, the plasma
component tends to be stable, because its low density implies low
kinetic energy. The neutral component, on the other hand, is
intrinsically unstable because it is not directly acted upon by
magnetic forces. Friction between the ions and neutrals allows
magnetic forces to act on the neutrals. As we will see, however, no
matter how large the magnetic field is, it cannot stabilize the
neutrals completely.

In the present discussion, we follow the analysis of Chandrasekhar
(1961) but consider two fluids, charged and neutral, coupled by
elastic collisions\footnote{We ignore coupling by ionization and
  recombination, since these processes are slow in comparison to
  elastic processes.}.
We assume the medium is incompressible ($\nabla\cdot{\bf v}_{i,n}$ =
0). Our problem is described by the neutral and ion momentum equations
and the magnetic induction equation:

\begin{eqnarray}
\rho_n\partial _t{\bf v}_n+\rho_n{\bf v}_n\cdot\nabla{\bf v}_n &=
  &-\nabla P_n-\rho_n\nu_{ni}({\bf v}_n-{\bf v}_i)\\
\rho_i\partial _t{\bf v}_i+\rho_i{\bf v}_i\cdot\nabla{\bf v}_i &=
  &-\nabla P_i-\rho_n\nu_{ni}({\bf v}_i-{\bf
    v}_n)+\frac{1}{4\pi}(\nabla\times{\bf B})\times{\bf
  B}
\end{eqnarray}
\begin{eqnarray}
\partial_t{\bf B} &= &\nabla \times ({\bf v_i \times B}),
\end{eqnarray}
where
\begin{eqnarray}
\nu_{ni} \equiv \gamma\rho_i
\end{eqnarray}
is the neutral-ion collision frequency, and
\begin{eqnarray}
\gamma \equiv \frac{<\sigma v>}{m_i+m_n}
\end{eqnarray}
is the collision rate coefficient per unit mass.

The system consisting of two constant density, constant pressure
regions of uniform flow aligned with a uniform magnetic field is
clearly a steady state solution of eqns. (1 - 3). 
We now consider small perturbations
of the flow. We assume these perturbations are independent of
$y$. According to Squires' Theorem (Drazin \& Reid 1981) and its
extension to MHD (Hughes \& Tobias 2001), in the absence of viscosity
and resistivity these are the fastest growing perturbations. Since the
magnetic field enters the problem only through the tension force, which is
independent of the structure in $y$, ion-neutral friction should not affect
this result. We
expand the momentum and induction equations to first order in a small
parameter $\epsilon$, writing
\begin{eqnarray}
v_x \equiv u&=&U+\epsilon u\\ v_y \equiv w&=&\epsilon w\\
B_x&=&B+\epsilon b_x\\ B_z&=&\epsilon b_z.
\end{eqnarray}
We assume all perturbed quantities $q$ take the form
\begin{eqnarray}
q&=&q_0e^{i(\omega t+k_xx)+\kappa z},
\end{eqnarray}
where $q_0$ is a constant amplitude. Linearizing in $\epsilon$ gives
\begin{eqnarray}
\rho_n(\partial_t +U\partial_x)u_n &= &-\partial_xP_n + \rho_n\nu_{ni}(u_i-u_n)\\
\rho_n(\partial_t +U\partial_x)w_n &= &-\partial_xP_n + \rho_n\nu_{ni}(w_i-w_n)\\
\rho_i(\partial_t +U\partial_x)u_i &= &-\partial_xP_i + \rho_n\nu_{ni}(u_n-u_i)\\
\rho_i(\partial_t +U\partial_x)w_i &= &-\partial_xP_i +
\rho_n\nu_{ni}(w_n-w_i) + \frac{B}{4\pi}(\partial_xb_z-\partial_zb_x)\\
(\partial_t + U\partial_x)b_x &= &B\partial_xu_i\\
(\partial_t + U\partial_x)b_z &= &B\partial_xw_i,
\end{eqnarray}
which correspond to the equations (180-183) in Ch. 12 of Chandrasekhar
(p. 508-9). Eliminating $B$ in equation (14) using equations (15) and (16),
constructing the $\hat z$ component of the curl of the momentum equations,
and applying $\nabla \cdot {\bf v}_{i,n} \equiv 0$, we obtain
\begin{eqnarray}
\left(\rho_i(\omega +Uk_x)-\frac{B^2k^2_x}{4\pi (\omega
  +Uk_x)}\right)(\partial^2_z -k^2_x)w_i &=
  &i\rho_n\nu_{ni}(\partial^2_z -k^2_x)(w_i -w_n)\\
\rho_n(\omega +Uk_x)(\partial^2_z -k^2_x)w_n &=
  &i\rho_n\nu_{ni}(\partial^2_z -k^2_x)(w_n -w_i)
\end{eqnarray}
These relations hold in each region, but not at the interface, where
the velocity is not analytic. Thus, to derive the dispersion relation,
we must integrate each equation over an infinitesimal region
surrounding $z=0$ and use standard Gaussian pillbox arguments. Denoting 
the jumps in quantities by $\Delta\{\}$, we get
\begin{eqnarray}\label{equ:jump}
\Delta\{\rho_i (\omega +Uk_x)\partial_zw_i\} &=
&i\Delta\{\rho_n\nu_{ni}\partial_z(w_i-w_n)\}+
\frac{B^2k^2_x}{4\pi}\Delta\{\frac{\partial_zw_i}{\omega +Uk_x}\}\\
\Delta\{\rho_n (\omega +Uk_x)\partial_zw_n\} &=
&i\Delta\{\rho_n\nu_{ni}\partial_z(w_n-w_i)\}.
\end{eqnarray}

We write the ion and neutral velocity amplitudes
in the form\\
\parbox{14cm}{
\begin{eqnarray*}\label{equ:zflows} 
w_{i1} &= &{\mathcal{A}}(\omega+U_1k)e^{kz} \\
w_{i2} &= &{\mathcal{A}}(\omega+U_2k)e^{-kz}\\
w_{n1} &= &{\mathcal{B}}(\omega+U_1k)e^{kz}\\
w_{n1} &= &{\mathcal{B}}(\omega+U_2k)e^{-kz}
\end{eqnarray*}}
\hfill \parbox{2.5cm}{
\begin{eqnarray}
z<0\\ 
z>0\\ 
z<0\\ 
z>0,
\end{eqnarray}}
where the subscripts $1$ and $2$ refer to the jet and ambient medium,
respectively. The forms of $w_{i,n}$ are chosen such that the
ion and neutral displacements are continuous across the
interface, as must hold on physical grounds. Substituting these forms into equations (19) and (20), we solve
for $\mathcal{B}$, the neutral velocity fluctuations, in terms of
$\mathcal{A}$, the ion velocity fluctuations. The result is
\begin{eqnarray}\label{equ:BA}
{\mathcal{B}}
&=&-i\frac{\rho_{n1}\nu_1(\omega+U_1k_x)+\rho_{n2}\nu_2(\omega+U_2k_x)}
{\rho_{n1}(\omega+U_1k_x)(\omega+U_1k_x-i\nu_1)+\rho_{n2}(\omega+U_2k_x)(\omega+U_2k_x-i\nu_2)}{\mathcal{A}}.
\end{eqnarray}
We then use equations (\ref{equ:zflows}) and (\ref{equ:BA}) in equation (\ref{equ:jump})
to obtain the dispersion relation
\begin{eqnarray}\label{equ:gendr}
&&\rho_{i1} (\omega + U_1 k)^2+\rho_{i2} (\omega +U_2k)^2 = (\rho_{n1}\nu_1(\omega +U_1k)+\rho_{n2}\nu_1(\omega +U_2k))\times\\
&&\left(i-\frac{(\rho_{n1}\nu_1(\omega +U_1k)+\rho_{n2}\nu_1(\omega +U_2k))}
{\rho_{n1}(\omega +U_1k)(\omega +U_1k-i\nu_1)+\rho_{n2}(\omega +U_2k)(\omega +U_2k-i\nu_2)}\right)+\nonumber\\
&&+\frac{B^2k^2}{2\pi}\nonumber
\end{eqnarray}
where $\nu_{ni}$ has been abbreviated to $\nu$ and $k_x$ abbreviated
to k.

We now assume that the jet and ambient medium have the same density,
work in the center of momentum frame so that
\begin{eqnarray}\label{equ:specialized}
U_1&=-U_2=&U\\
\rho_{n1}&=\rho_{n2}=&\rho_n\nonumber\\
\rho_{i1}&=\rho_{i2}=&\rho_i,\nonumber
\end{eqnarray}

and nondimensionalize the problem as follows:
\begin{eqnarray}\label{equ:dimless}
m&\equiv&\frac{\rho_n}{\rho_i}\\
x&\equiv&\frac{\omega}{\nu}\nonumber\\
h&\equiv&\frac{Uk}{\nu}\nonumber\\
a&\equiv&\frac{\sqrt{m} Bk}{\nu\sqrt{4\pi\rho_n} } = \frac{kv_{Ai}}{\nu},\nonumber
\end{eqnarray}
where the ion Alfv\'en speed $v_{Ai}$ is defined as $
\frac{B}{\sqrt{4\pi\rho_i}}$. The variables $x$, $h$ and
$a$ represent the perturbation frequency, flow speed and ion Alfv\'en
speed, respectively.

Substituting equations (\ref{equ:specialized}) and (\ref{equ:dimless})
into equation (\ref{equ:gendr}) we obtain
\begin{eqnarray}\label{equ:drfull}
x^4-i(m+1)x^3+(2h^2-a^2)x^2-i(mh^2-a^2+h^2)x+h^2(h^2-a^2) &=&0.
\end{eqnarray}

In all cases of interest, $m\gg 1$ and $a\gg h$, which allows us to
drop a few of the terms in equation (\ref{equ:drfull}). This leads to the
somewhat more compact expression
\begin{eqnarray}\label{equ:drfinal}
x^4-imx^3-a^2x^2-i(mh^2-a^2)x-a^2h^2&=&0.
\end{eqnarray}
Equation (\ref{equ:drfinal}) is the basis of our subsequent stability
analysis.  We will be looking for solutions for $x$ which have a
negative imaginary component, an indication of a growing instability.

\section{Analysis of the Dispersion Relation}

Equations (\ref{equ:drfull}) and (\ref{equ:drfinal}) reduce, in special cases,
to previously known results. These are discussed in
\S\ref{subsec:recovery}.
In \S\ref{subsec:modes} we present numerical and analytical solutions
for the fastest growing modes. In \S\ref{subsec:eigenfunctions} we
discuss the physical nature of the instabilities.

\subsection{Recovery of previously known results}\label{subsec:recovery}

If $h\equiv 0$, equation (\ref{equ:drfinal}) reverts to the dispersion
relation for Alfv\'en waves in a partially ionized medium (Kulsrud \&
Pearce 1969)
\begin{equation}\label{equ:hmdr}
x\left(x^2-a^2\right)-i\left(x^2-\frac{a^2}{m}\right)=0.
\end{equation}
At sufficiently small wavenumber ($k< \frac{2\nu_{ni}}{v_{An}}$;
$a<2\sqrt{m}$) the wave speed is close to the neutral Alfv\'{e}n
velocity $v_{An}$ because the neutral-ion collision timescale is less
than the wave period and the ion and neutral species are closely
coupled. We can derive this result from equation (\ref{equ:hmdr}) by
treating the second factor in parentheses as dominant. This leads to
\begin{eqnarray}
x\approx
\pm\frac{a}{\sqrt{m}} + i\frac{a^2}{2m},
\end{eqnarray}
which is a weakly damped Alfv\'en wave propagating at speed v$_{An}$.

At sufficiently large wavenumber ($k > \frac{\nu_{in}}{2 v_{Ai}}$;
$a>m/2$), the wave period is shorter than the ion-neutral collision
timescale. Thus, Alfv\'{e}nic disturbances are able to propagate
through the ions without disturbing the neutrals and propagate at the
ion Alfv\'{e}n velocity $v_{Ai}$. We can derive this result from equation
(\ref{equ:hmdr}) by treating the first factor in parentheses as
dominant. This leads to
\begin{eqnarray}
x\approx\pm~a + i\frac{m}{2},
\end{eqnarray}
which is a damped wave propagating at the ion Alfv\'en velocity. In
the intermediate wavenumber regime ($2\sqrt{m}<a<m/2$), waves cannot
propagate because the ions are poorly coupled to neutrals and thus
feel a strong drag.

The shear instability in a fully-neutral medium (or ionized but
magnetic field free) is present at any velocity and has growth rate
$kU$ ($\equiv h$ in our notation).  In order to recover the
hydrodynamic instability we must use equation (\ref{equ:drfull}) with
$a\equiv 0$. The dispersion relation can then be written in the form
\begin{equation}\label{equ:drhydro}
\left(x^2+h^2\right)\left[x^2-i\left(m+1\right)x+h^2\right]=0.
\end{equation}
The hydrodynamic instability is the root $x=-ih$. 

On the other hand, a fully ionized medium with a magnetic field
parallel to the shear flow is stabilized by magnetic tension if $U <
v_A$. In order to recover the MHD instability, we set $m\equiv 0$ in
equation  (\ref{equ:drfull}) and write the dispersion relation in the form
\begin{equation}\label{equ:drplasma}
\left(x^2+h^2-a^2\right)\left(x^2+h^2-ix\right)=0.
\end{equation}

Note that the growth rates of both hydrodynamic and MHD instabilities
increase linearly with wavenumber.
\subsection{The most unstable roots}\label{subsec:modes}

With \S\ref{subsec:recovery} as background, we now turn to the shear
instability in the partially ionized regime.

If there were no ion-neutral friction, the neutrals would always be
unstable. The plasma would be stable if $U<v_{Ai}$ ($h<a$), which we
assume to be the case. Since the growth rate of the hydrodynamic
instability increases with $k$, there is a wavenumber above which
friction with the ions cannot stabilize the perturbation.  Therefore,
we expect short wavelength perturbations to grow at the hydrodynamic
rate in the neutral fluid, but to leave the plasma and magnetic field
in place.

At longer wavelengths, there is time for ion-neutral friction to
act. If $v_A<U$ ($a<h\sqrt{m}$), the magnetic field cannot be expected
to stabilize the medium no matter how well the ions and neutrals are
coupled. On the other hand, if $U<v_A$, we might expect that at
sufficiently long wavelengths the coupling is good enough to
completely stabilize the system. We will see, however, that this is
never the case.

We carried out a parameter study of the roots of equation
(\ref{equ:drfinal}) as functions of $a$ and $h$ using the Mathematica
software package (Wolfram 1999). Surprisingly, the roots are closely
approximated by the analytical expressions given in Table~\ref{roots}
for the specified range of dimensionless parameters. In the results
presented here, we focus on two sets of physical environments which we
believe are illustrative of the range of conditions encountered in
molecular outflows. The density of the ambient medium near massive
protostars is a strong function of distance from the protostar.
within radii of 100-1000 AU densities are >10$^5$ cm$^{-3}$, further
out where outflows can be easily resolved and structure studied
densities of 10$^3$ cm$^{-3}$ are typical. In both examples, we take
the molecular hydrogen density $n(H_2)$ to be 10$^3$ cm$^{-3}$, the
relative flow speed $2U$ to be 20 km s$^{-1}$, and the ratio $m$ of
neutral to ionized mass density to be 10$^6$ (corresponding to an
ionization fraction $\sim 10^{-7}$ for an ion to neutral particle mass
ratio $\sim 10$). The collision frequency $\nu$ is $1.5\times
10^{-13}$ s$^{-1}$ (we use the rate coefficient of Draine, Roberge, \&
Dalgarno (1983); $\langle\sigma v\rangle = 1.5\times 10^{-9}$ cm$^2$
s$^{-1}$).  With the these values, the wavelength at which the
hydrodynamic frequency $kU$ matches the collison frequency ($h=1$) is
about 1.3 pc.  The only difference between the two cases is in the
magnetic fieldstrength, which we take to be 1 $mG$ (strong field case)
and 1 $\mu G$ (weak field case). With these parameters, $a/h = 4400$
and $4.4$ in the strong and weak field cases, respectively.

\begin{figure}
\includegraphics[angle=90,height=12cm]{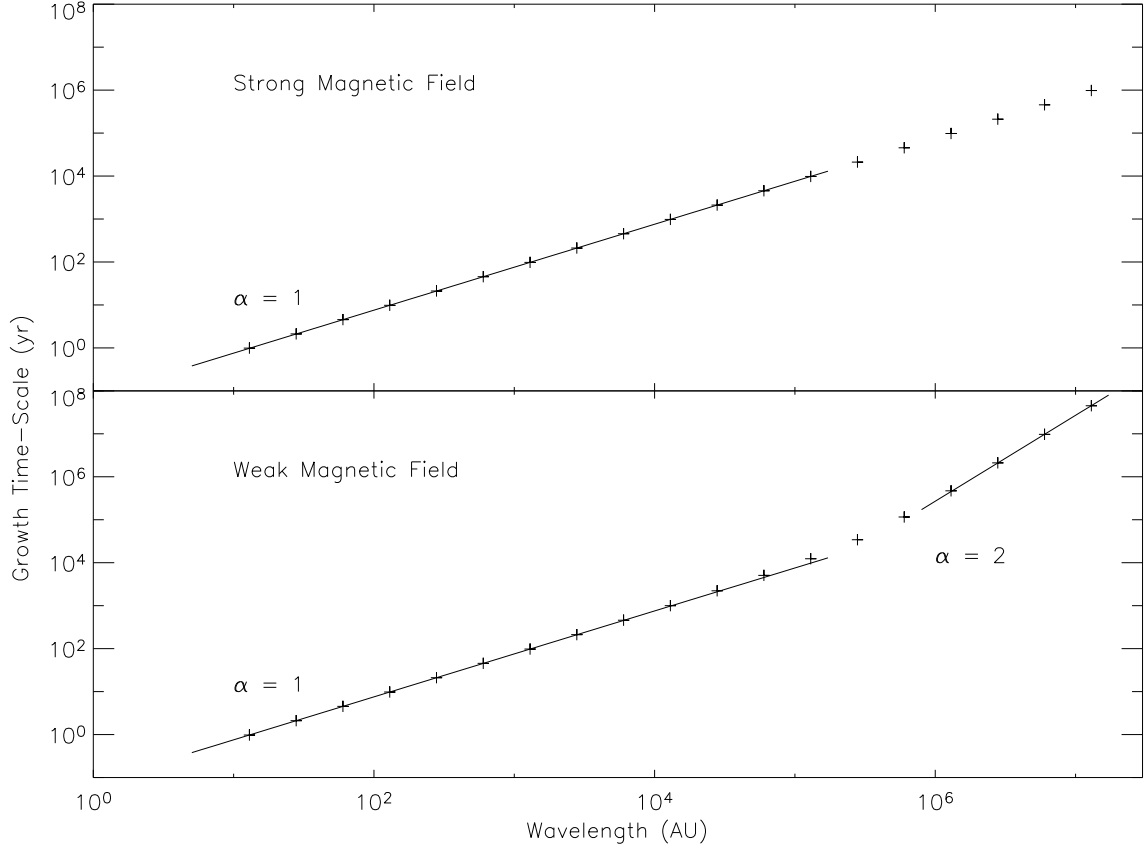}
\caption{The growing instability root of the characteristic equation
  as a function of wavelength. The strong magnetic field regime (top)
  assumes $U=$10 km s$^{-1}$ and $B=$1mG and the weak magnetic field
  regime (bottom) assumes $U=$10 km s$^{-1}$ and $B=$1$\mu$G. The
  crosses represents the numerically determined root. The solid lines
  represent the analytic approximation to root, $h$ ($h$) at short
  wavelengths and $h$ ($h^2$) at long wavelengths for the strong
  (weak) magnetic field regime. \label{growth}}
\end{figure}

Since only growing modes will be dynamically important in the
interface between a fast moving jet and the ambient ISM, we restrict
our discussion to two solutions, $-ih$ [and the variant
$-i\left(\frac{1}{2} - h\right)$] and $-ih^2$. Confirmation that these
expressions closely approximate the numerically derived roots is shown
in Figure \ref{growth}. The numerically determined roots to the
dispersion equation are shown as a function of wavelength; overplotted
are the analytical approximations to the two growing modes. The
wavelengths $2\pi/k$ and growth times $(Im(\omega))^{-1}$ are plotted
in physical units for the two cases specified above.

The analytical solutions can be derived from the dispersion relation
by dropping small terms. The dominant terms used to derive each
solution are given in Table~\ref{roots}. For example, the first root
listed, $-ih$, is the standard hydrodynamic instability derived from
equation  (\ref{equ:drhydro}). From the dispersion relation, we can see
that the terms $\varpropto a^2$ ($-a^2x^2$, $ia^2x$, $-a^2h^2$)
dominant in the high magnetic field regime. Of the $a^2$ terms, the
terms $\varpropto h^2$, $x^2$ dominate over the term $\varpropto x$
because $h > 1$. In contrast, when $h < 1$, the terms $\varpropto x$,
$h^2$ dominate, and the growing mode is $-i h^2$.

\begin{figure}
\includegraphics[angle=90,height=12cm]{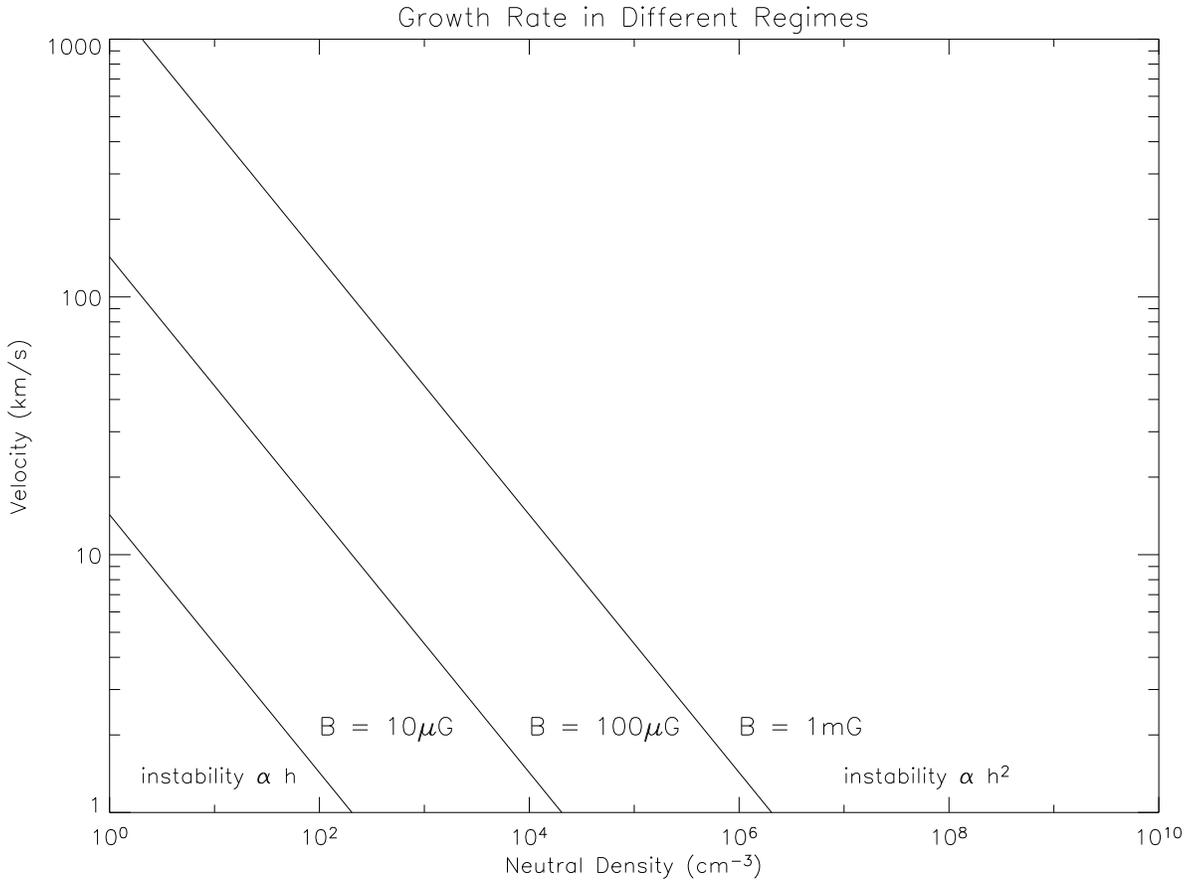}
\caption{The growth rate of the Kelvin-Helmholtz instability depends
  on the density, flow velocity and magnetic field strength. The
  transition between the standard Kelvin-Helmholtz instability (growth
  rate $\propto$ h) and the new, slower instability (growth rate
  $\propto$ h$^2$) is given for 3 magnetic field strengths.
   \label{regime}}
\end{figure}

This second growing mode, $-ih^2$, is a new mode that arises only in a
weakly ionized medium. It originates when $h<1$, or equivalently $kU
<\nu$. The neutrals feel a constant drag force, independent of
position, caused by collisions with the stationary ions. The
instability has an origin similar to the standard hydrodynamic
instability explained in \S 2. The principal difference is that the
neutral-ion collision time is shorter than the wave period. Thus,
neutral material collides with ions (held stationary because of the
magnetic field tension) frequently during each perturbation period,
which slows the growth of small perturbations. This slowing of the
growth is clearly visible in Figure~\ref{growth} (bottom panel) where
the growth time is longer for long wavelength (and long period)
perturbations. The origin of this mode can be made more clear by
assuming the ions are held strictly in place by the magnetic field and
re-deriving the roots (see Appendix A). The break between the two
growing modes occurs when $h=1$ or, equivalently, $\frac{1}{Uk}$ (the
wave period) = $\frac{1}{\nu}$ (the neutral-ion collision timescale).
See Figure~\ref{regime} for physical parameters of this transition.
Since this mode grows more slowly than the $-ih$ mode, it is important
only in the situation $U<v_A$ (or $\sqrt{m}h <a $), when the MHD mode
is stable.

When  $\sqrt{m}h > a $ and $h<1$, $-ih\sqrt{1-a^2/mh^2}$ is a better approximation
to the growth rate. This is the analog of eqn. (\ref{equ:drplasma}) for a fully ionized medium.
It is based on the limit of infinitely strong ion-neutral coupling.

\subsection{Physical nature of the unstable modes}\label{subsec:eigenfunctions}

One of the main results of \S\ref{subsec:modes} is that the plasma and
the neutrals can follow very different dynamics as the instability
grows. Although the interface between the two regions of flow is the
same for the plasma as for the neutrals in the equilibrium system, the
displaced interfaces for the plasma and neutral species can
differ, as can their
velocities. The negligibly small inertia of the ions renders any differences
irrelevant to the basic mechanism of instability, but they are important for
the nonlinear outcome of the instability, and for its spectroscopic
signatures.

On physical grounds, we expect large differences between $v_i$ and $v_n$
when ion - neutral coupling is weak ($h>1$) and magnetic forces are strong.
Whether these differences persist at small $h$ depends on the relative values
of $a/\sqrt{m}$ and $h$. If their ratio is less than unity the field is too
weak to suppress instability even in the limit of infinitely strong ion-neutral
coupling, and we expect $v_i\sim v_n$. If $a/h\sqrt{m} > 1$ the instability
acts on the neutrals but leaves the ions behind, and we expect large
differences between $v_i$ and $v_n$ even at short wavelengths. 

\begin{figure}
\includegraphics[angle=90,height=12cm]{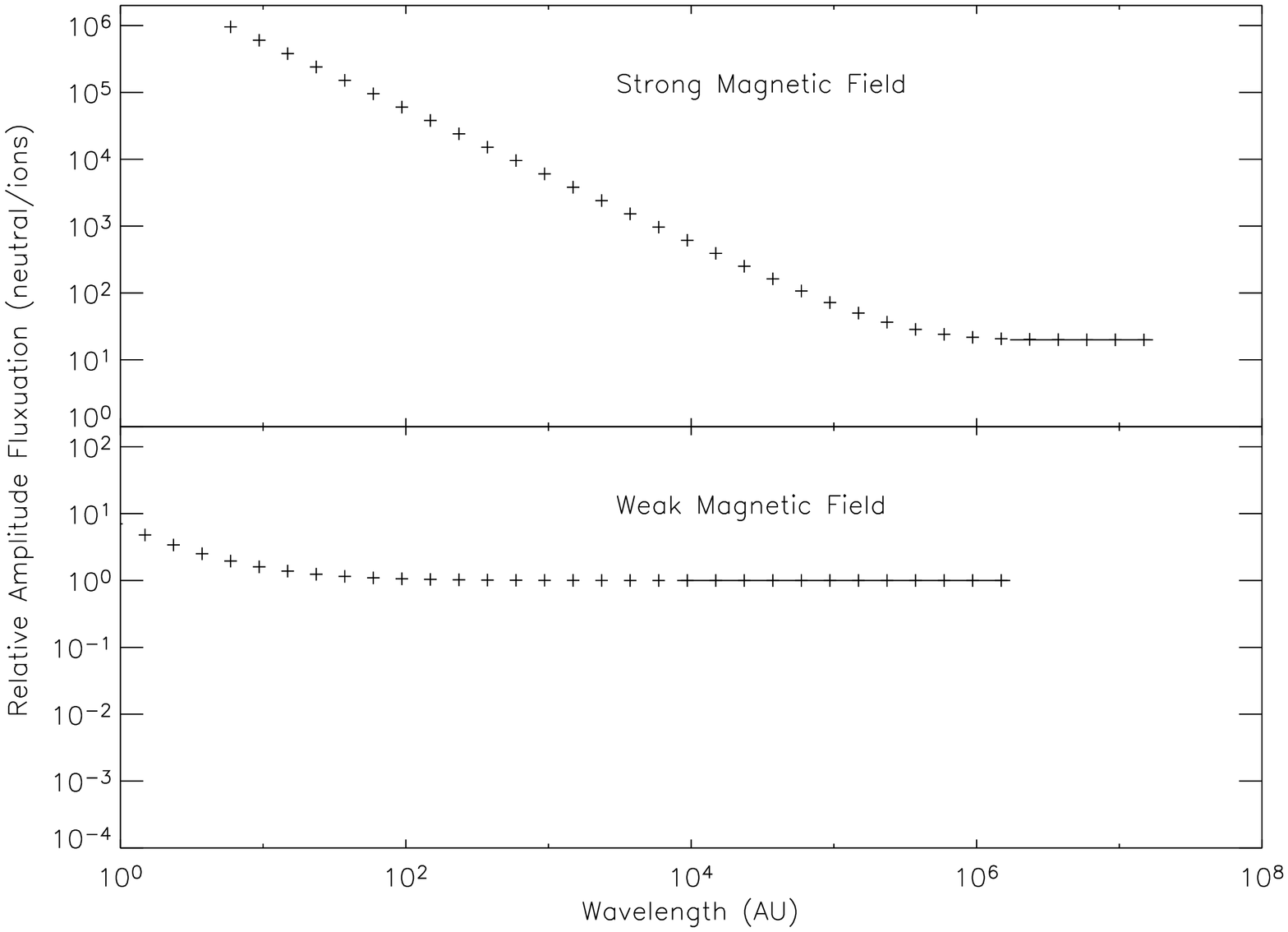}
\caption{The amplitude of the neutral perturbations relative to the
  ion perturbations. Numerical results are represented by crosses. The
  lines represents the analytic approximations to the relative
  perturbations as specified above. The strong magnetic field regime
  (top) assumes $U=$10 km s$^{-1}$ and $B=$ 1 mG and the weak magnetic
  field regime (bottom) assumes $U=$10 km s$^{-1}$ and
  $B=$ 1 $\mu$G. \label{amp}}
\end{figure}

These expectations are borne out by analysis.
According to equation (\ref{equ:zflows}), the amplitudes of the ion and
neutral displacements are $\mathcal{A}$ and $\mathcal{B}$,
respectively. Their ratio follows from equation  (\ref{equ:BA}), which for
the problem at hand can be written in dimensionless form using
equations (\ref{equ:dimless}) as
\begin{eqnarray}\label{equ:BAratio}
\frac{{\mathcal{B}}}{{\mathcal{A}}} &= &\frac{-ix}{x^2+h^2-ix}.
\end{eqnarray}
The modulus of the relative amplitude is plotted as functions of wavelength in
Figure \ref{amp} for the $v_A>U$ and $v_A<U$ cases, respectively. When the
field is strong and $h$ is large (short wavelength), the ions are held firmly in place
and the instability is almost entirely in the neutrals, so
$\vert{\mathcal{B}}/{\mathcal{A}}\vert$ is large. The numerical results are well fit by the
asymptotic formula $\vert{\mathcal{B}}/{\mathcal{A}}\vert\sim a^2/mh\propto\lambda^{-1}$. As $h$
decreases the coupling improves, but the magnetic restoring force is still very large and the ion response 
never matches the neutrals. In this regime,  $\vert{\mathcal{B}}/{\mathcal{A}}\vert\sim a^2/mh^2$. Recall (see
Figure 2) that in this limit the instability is of the weak form $x\sim -ih^2$.

When the field is weak, magnetic forces are relatively unimportant. In
this case, $\vert{\mathcal{B}}/{\mathcal{A}}\vert$ is nearly unity
over most of the range considered, although Figure \ref{amp} (bottom
panel) shows a slight upturn at the shortest wavelengths, at which
ion-neutral friction is relatively weak. The location of the upturn
shows the expected scaling with $B$. For example, at $\lambda=280$,
$\vert{\mathcal{B}}/{\mathcal{A}}\vert \sim 1$ for $B=1\mu G$,
increasing to 25 if $B=10\mu G$ and 50 if $B=20 \mu G$.  In all three
cases, however, $\vert{\mathcal{B}}/{\mathcal{A}}\vert$ drops to unity
for $h <$ 1 ($\lambda\sim 2.8\times 10^6$ AU). Not until the field is
strong enough to stabilize the bulk medium $B\sim 230\mu G$ does
$\vert{\mathcal{B}}/{\mathcal{A}}\vert$ exceed unity for $h < 1$.

In deriving both the numerical and asymptotic results 
in the cases where $x\sim -ih$,  the $x^2$ and $h^2$ terms
in the denominator of eqn. (\ref{equ:BAratio}) nearly
cancel. Thus, the right hand side of the equation (\ref{equ:BAratio}) must be computed using
roots to the exact dispersion relation (equation
(\ref{equ:drfull})). 

The critical result from this analysis is that if $B\ge\sim 10\mu G$
then at relevant length
scales and timescales for bipolar outflows ($l <$ 1000 AU and $\tau <$
1000 years), the fluctuations are primarily in the neutrals and not in
the ions or magnetic field. In contrast to the MHD shear flow
instabilities simulated by Jones et al. (1996) and Malagoli et
al. (1996), the magnetic field cannot wind up and dissipate on small
scales, and the instability is primarily hydrodynamic in character.
We now turn to the observational implications of this conclusion.

\section{Observational Implications}

The calculations presented here do not permit quantitative
predictions for observational signatures of the 
two instabilities for ambient ISM entrainment by massive bipolar
outflows.  We would need to develop numerical models 
with two fluid MHD that incorporate
non-planar geometry and non-linear
growth, and would need chemical models and  radiative transfer calculations to predict the
resulting spectra.
Such an analysis is beyond the scope of this paper. However,
several physical signatures should be robust, and may be observable.

\subsection{Molecular Spectroscopy} 

As we showed in \S 3.3, unless $B$ is less than about 10 $\mu$G, over most of
the parameter space of interest
only neutrals participate in the resulting
turbulence. That is, only neutrals will have a velocity significantly
different from either the outflow jet or the ambient ISM. For example,
if outflow jets had uniform velocity emerging from the protostar, then
the emission from ionized molecules should not be detectable at
velocities different from the outflow velocity. This signature should
be particularly distinctive if there are ion species that are present
in only one of these media. Outflow jets probably have a range in
velocity, however, which will make isolating a velocity unique to the
interface region difficult.

If the transition between the jet and the ambient medium can be
resolved, we might expect relatively large differences between the
profiles of charged and neutral species. The ion flow would be
laminar, and would show a smooth transition between the velocity of
the core of the jet and the velocity of the ambient medium. The
neutral flow would show the same range of velocities, but would be
turbulent, leading to broader lines.

\subsection{Heating}

Figure \ref{amp} predict that throughout most of parameter space, the
magnetic field prevents the ions from participating in unstable
perturbations.  The resulting velocity difference between the ions and
neutrals creates friction that can heat the gas. The heating rate per
unit volume is given by
\begin{eqnarray}
\Gamma_{fric} &= &\rho_n\nu_{ni}(v_n-v_i)^2\approx\rho_n\nu_{ni}v_n^2.
\end{eqnarray}
For example, assuming $v_n\sim 1$ km s$^{-1}$, $n_n=10^3$ cm$^{-3}$,
$n_i=$ 10$^{-4}$ cm$^{-3}$ gives $\Gamma_{fric}\sim 6\times 10^{-24}$
ergs cm$^{-3}$ s$^{-1}$. Assuming that the heating is balanced by
radiative cooling, the corresponding luminosity density is 0.04
$L_{\odot}$ pc$^{-3}$.  The frictional heating rate dominates cosmic
ray heating.  Taking the latter from Spitzer (1978) we find
\begin{equation}\label{equ:heatingratio}
\frac{\Gamma_{fric}}{\Gamma_{CR}}=1.5\times 10^6n_i v_5^2,
\end{equation}
where $v_5$ is the neutral velocity in km/s. Unlike viscous heating,
which is also strong in a turbulent boundary layer, $\Gamma_{fric}$ is
independent of the thickness of the layer.

\subsection{Magnetic field configuration}

The orientation of the magnetic field in the outflow could be mapped
by measuring the polarization of infrared radiation emitted by the
dust grains. According to our analysis, the magnetic field should
remain well organized even though the neutral flow is turbulent,
provided that the mean field is aligned with the flow (if the field
were transverse to the flow, the plasma would be unstable, and would
carry the field with it, but the turbulent motions would be
perpendicular to the field, and would not strongly bend it).

Observations of a turbulent field in the outflow could be explained if
the ionization fraction were much larger than assumed here, resulting
in stronger coupling, or if the grains were aligned by the turbulent
neutral flow rather than by the magnetic field. Observations of a
straight field, however, demonstrate only that the field is well
organized at or above the scale resolved by the observations
(Heitsch et al. 2001). A low
polarized intensity could be evidence for a disordered field below the
resolved scale, but could also be due to poor alignment of the grains
(Padoan et al. 2001).

\section{Conclusions}

Outflows from massive protostars are observed to carry a large mass
flux and to be poorly collimated. Both properties suggest that these
outflows entrain substantial volumes of ambient ISM
material. Turbulence driven by shear flow instabilities at the
interface between the jet and the surrounding medium is one possible
mechanism for entrainment.

In this paper we have examined the Kelvin-Helmholtz instability in a
weakly ionized medium with parameters chosen to be similar to
molecular outflows.  In order to study effects introduced by two
fluids (plasma and neutral), we chose the simplest possible geometry:
two uniform media (taken in all examples to have the same density) in
relative motion at speed $U$, separated by a sharp interface, with a
uniform magnetic field parallel to the flow. Under these conditions, a
plasma is unstable if the relative velocity exceeds the Alfv\'en speed
(Chandrasekhar 1961). For the parameters of interest to our problem,
the plasma is magnetically stabilized while the neutrals, in the
absence of frictional coupling to the ions, would always be
unstable. In the limit of perfect frictional coupling, the combined
system - with the Alfv\'en speed based on the total mass density
instead of the plasma density - could be either stable or unstable, according
to whether the bulk Alfv\'en speed $v_A$ exceeds $U$.

Most of the analysis in our paper is based on the dispersion relation
derived in \S 2, equation (\ref{equ:drfinal}).  Our main result is
that for much of the size range of interest, the ions and neutrals are
sufficently decoupled that the neutrals are unstable, while the ions
are held in place by the magnetic field. At short scales, the growth
rate is well approximated by the growth rate of the hydrodynamic
instability, $kU$. At longer scales, the fastest growing mode is
either the magnetically modified MHD mode, if the magnetic field is
weak enough ($v_A < U$), or, if $v_A > U$, a new mode with growth rate
$kU(kU/\nu)$. The transition between these modes of instability occurs
at $kU/\nu \approx 1$. The dispersion diagram is plotted in Figure
\ref{growth}. The relative amplitudes of the ion and neutral
fluctuations are shown in Figure \ref{amp}, demonstrating
that over most of the regime of interest, the ions drop out of the
instability.

We believe these results continue to hold in cylindrical geometry. In
fact, they should be even more pronounced because at short wavelengths
the cylindrical and planar cases coincide, while at long wavelengths
the instability is suppressed by geometry. We also expect that more
general perturbations, such as those with a large transverse
wavenumber $k_y$, would be characterized by similar ion-neutral
decoupling.

Although we have not followed the instability into the nonlinear
stage, there is no reason why frictional coupling should be more
efficient at large amplitude, so we expect our results to apply to the
nonlinear regime. There is, however, a compelling case to be made for
nonlinear calculations: they are necessary to produce detailed
observational predictions. As such, they should follow the thermal and
chemical state of the gas, as well as its dynamics.

Based on the linear theory, however, it appears that our model may be
tested in several ways, mentioned in \S 4; differences in line
profiles between charged and neutral species, efficient heating by
ion-neutral friction, and the polarimetric signature of a well
organized magnetic field. We will report on spectroscopic evidence in
a forthcoming paper (Watson et al. 2004).

Our results do, however, leave us with a conundrum. Churchwell (1997)
pointed out that the rate of entrainment required to explain the mass
flux in outflows is unreasonably large, if the entrainment is
hydrodynamic in nature. The low degree of ion-neutral coupling
predicted by our model suggests that the entrainment is more
hydrodynamic than magnetohydrodynamic, although nonlinear simulations
and analysis are necessary to settle the issue.

\acknowledgements 

We are happy to acknowledge support by NSF grant AST-0328821, NSF
PHY-0215581 and by the Graduate School of the University of Wisconsin,
Madison. F.H. is grateful for support by a Feodor-Lynen Fellowship
from the Alexander von Humboldt Foundation.  \appendix
\section{Ions Held Stationary}
We begin with the same assumptions as specified in section 2, with the
additional assumption that the ions remain at rest. That is, they are
not disturbed by the perturbation. The equation of motion for the
neutral particles is:
\begin{eqnarray}
\rho_n\partial _t{\bf v}_n+\rho_n{\bf v}_n\cdot\nabla{\bf v}_n &=
  &-\nabla P_n-\rho_n\nu_{ni}{\bf v}_n
\end{eqnarray}
Assuming a periodic perturbation, we obtain:
\begin{eqnarray}
\rho_n (\omega +Uk_x)u_n &= &-k_xP_n + i\rho_n \nu u_n\\
\rho_n (\omega +Uk_x)w_n &= &i\partial_z P_n + i\rho_n \nu w_n.
\end{eqnarray}
Taking the curl and combining to eliminate the pressure term, we
obtain
\begin{eqnarray}
\rho_n(\omega +Uk_x-i\nu)\left(\partial^2_z -k_x^2\right)w_n &= &0
\end{eqnarray}
We integrate over the interface to obtain,
\begin{eqnarray}
\Delta\{\rho_n(\omega+Uk)\partial_z
w_n\}&=&i\Delta\{\rho_n\nu\partial_z w_n\}
\end{eqnarray}
We introduce a solution of the form

\parbox{10cm}{
\begin{eqnarray*}
w_{n1}&=&\mathcal{A}(\omega+Uk)e^{kz}\\
w_{n2}&=&\mathcal{A}(\omega+Uk)e^{-kz}
\end{eqnarray*}}
\hfill \parbox{3cm}{
\begin{eqnarray}
z<0\\
z>0
\end{eqnarray}}

to obtain
\begin{eqnarray}
\rho_n(\omega+Uk)^2-i\rho_n\nu(\omega-Uk)&=&0
\end{eqnarray}
After converting to dimensionless notation, we obtain
\begin{eqnarray}
x^2+h^2-ix &= &0\\
x&=&\frac{1}{2}\left( i\pm\sqrt{-1-4h^2}\right)\\
x&\approx &\frac{1}{2}\left( i\pm i\left( 1+2h^2\right)\right)
\end{eqnarray}
For h$\ll$1,
\begin{eqnarray}
x&\approx &i,-ih^2.
\end{eqnarray}
Thus, we see that the $-ih^2$ mode observed in section 2 originates
from the uniform drag that the neutral particles feel from the
uniformily distributed ionized particles.


\begin{deluxetable}{rllr}
\tablecaption{Unstable Roots of Characteristic Equation \label{roots}}
\tablehead{
&\multicolumn{3}{l}{$x^4-imx^3-a^2x^2-i(mh^2-a^2)x-a^2h^2=0$}\\
&\multicolumn{3}{l}{(1)\hspace{\spc} (2)\hspace{\spc} (3)\hspace{\spc}
  (4)\hspace{\spc} (5)\hspace{\spc} (6)}\\\hline\\
&\colhead{Root}
&\colhead{Dominant Terms} &\colhead{Physical Regime}}
\startdata
&$-ih$    &3, 6    &$h>1$\\
&$-ih^2$  &5, 6  &$1>h$ and $\frac{a}{\sqrt{m}}>h$\\
&$i(\frac{1}{2}\pm h)$   &3, 4, 5, 6 &$\frac{a}{\sqrt{m}}>h>1$\\
\enddata
\end{deluxetable}

\begin{thebibliography}{}
\bibitem[Bachiller \& Tafalla(1999)]{1999osps.conf..227B} Bachiller, R.~\& 
Tafalla, M.\ 1999, NATO ASIC Proc.~540: The Origin of Stars and Planetary 
Systems, 227
 
\bibitem[Batchelor (1967)]{} Batchelor, G.K. 1967, {\em An Introduction to
Fluid Dynamics}, Cambridge University Press

\bibitem[Chandrasekhar (1961)]{} Chandrasekhar, S. 1961, {\em Hydrodynamic and
Hydromagnetic Stability}, Oxford University Press

\bibitem[Churchwell(1997)]{1997ApJ...479L..59C} Churchwell, E.\ 1997, 
\apjl, 479, L59
 
\bibitem[Drazin \& Reid(1981)]{} Drazin, P.~G.~\& Reid, 
W.~H.\ 1981, {\em Hydrodynamic Stability}, Cambridge Univ. Press
 
\bibitem[Heitsch et al. (2001)]{2001ApJ...561..800H} Heitsch, F., Zweibel,
E.G., Mac Low, M-M., Li, P., \& Norman, M.L. 2001, \apj, 561, 800

\bibitem[Hughes \& Tobias (2001)]{} Hughes, D.W., \& Tobias, S.M. 2001,
Proc. Roy. Soc. Lond. A., 457, 1365

\bibitem[Jones, Gaalaas, Ryu, \& Frank(1997)]{1997ApJ...482..230J} Jones, 
T.~W., Gaalaas, J.~B., Ryu, D., \& Frank, A.\ 1997, \apj, 482, 230

\bibitem[Kulsrud \& Pearce(1969)]{1969ApJ...156..445K} Kulsrud, R.~\& 
Pearce, W.~P.\ 1969, \apj, 156, 445 

\bibitem[Malagoli, Bodo, \& Rosner(1996)]{1996ApJ...456..708M} Malagoli, 
A., Bodo, G., \& Rosner, R.\ 1996, \apj, 456, 708
 
\bibitem[Padoan et al. (2001)]{2001ApJ...559.1005P} Padoan, P., Goodman, A.,
Draine, B.T., Juvela, M., Nordlund, A., \& R\"ognvaldsson, \"O.E. 2001, \apj,
559, 1005

\bibitem[Spitzer(1978)]{1978ppim.book.....S} Spitzer, L.\ 1978,
  Physical Processes in the Interstellar Medium (New York:
  Wiley-Interscience)

\bibitem[Watson, Churchwell, Zweibel, \& Crutcher (2004)]{} Watson, C.,
  Churchwell, E., Zweibel, E.G., Crutcher, R. 2004, in prep.

\bibitem[Wolfram (1999)]{} Wolfram, S. 1999, The Mathematica Book (4th
ed.; Champaign/New York: Wolfram Media/Cambridge Univ. Press)

\end{thebibliography}
\end{document}